\documentclass[12pt,epsf,amsmath,amssymb,amsfonts]{article}
\usepackage{tabularx}
\usepackage{array}
\usepackage{graphics}
\usepackage{graphicx}
\usepackage{epsfig}
\usepackage{amsmath}
\usepackage{amssymb}
\usepackage{citesort}
\usepackage{color}
\makeatletter

\usepackage{verbatim}

\newcommand{\msbar}{{\overline{\rm MS}}}

\newcommand{\bea}{\begin{eqnarray}}
\newcommand{\eea}{\end{eqnarray}}
\newcommand{\beq}{\begin{equation}}
\newcommand{\eeq}{\end{equation}}

\newcommand{\pdir}{p\kern -5.2pt\raise 0.2ex\hbox {/}}
\newcommand{\vdir}{v\kern -5.75pt\raise 0.15ex\hbox {/}}
\newcommand{\kdir}{k\kern -5.75pt\raise 0.15ex\hbox {/}}
\newcommand{\epsdir}{\epsilon\kern -5.0pt\raise 0.15ex\hbox {/}}
\newcommand{\bvdir}{\bar{v}\kern -5.75pt\raise 0.15ex\hbox {/}}
\newcommand{\Ddir}{D\kern -7.75pt\raise 0.20ex\hbox {/}}
\newcommand{\ldir}{l\kern -5.0pt\raise 0.2ex\hbox{/}}
\newcommand{\varepsdir}{\varepsilon\kern -5.5pt\raise 0.15ex\hbox{/}}

\newcommand{\kkbar}{K^0-{\overline{K^0}}}

\def\la{\langle}
\def\ra{\rangle}

\def\l{\left}
\def\r{\right}
\def\nn{\nonumber}

\makeatother

\begin{document}

\thispagestyle{empty}
\begin{flushright}
\begin{tabular}{l}
{\tt  Orsay 03-79}\\
{\tt  Roma-1362/03}\\                                              
\end{tabular}
\end{flushright}
\vskip 2.8cm\par
\begin{center}
{\par\centering \textbf{\LARGE Impact of the finite volume effects} }\\
\vskip .25cm\par
{\par\centering \textbf{\LARGE on the chiral behavior of $f_K$ and $B_K$} }\\
\vskip 1.85cm\par
{\par\centering \large
\sc Damir~Be\'cirevi\'c$^a$ and Giovanni~Villadoro$^b$}
\end{center}
{\par\centering \textsl{$^a$Laboratoire de Physique Th\'eorique (B\^at.210), Universit\'e Paris Sud}\\
\textsl{Centre d'Orsay, F-91405 Orsay-Cedex, France.}\\
\vskip 0.3cm\par} 
{\par\centering \textsl{$^b$Dip.~di Fisica, Univ.~di Roma ``La
Sapienza''
and}\\
\textsl{INFN, Sezione di Roma, P.le A.~Moro 2, I-00185 Roma,
Italy}\textsf{\textit{.}}\\
\vskip 2.1cm\par }                                                               
%\hrule
\begin{abstract}
{We discuss the finite volume corrections to $f_K$ and $B_K$ 
by using the one-loop chiral perturbation theory in full, 
quenched, and partially quenched QCD. 
We show that the finite volume corrections to these quantities dominate  
the physical (infinite volume) chiral logarithms.}
\end{abstract}
\newpage

\setcounter{page}{1}
\setcounter{footnote}{0}
\setcounter{equation}{0}                                                        

%%%%%%%%%%%%%%%%%%%%%%%%%%%%%%%%%%%%%%%%
%%%%%%%%%%%%%%%%%%%%%%%%%%%%%%%%%%%%%%%%
\noindent
 
\renewcommand{\thefootnote}{\arabic{footnote}}
 
%\newpage
\setcounter{page}{1}
\setcounter{footnote}{0}
\setcounter{equation}{0}
%%%%%%%%%%%%%%%%%%%%%%%%%%%%%%%%%%%%%%%%
%%%%%%%%%%%%%%%%%%%%%%%%%%%%%%%%%%%%%%%%
%%%%%%%%%%%%%%%%%%%%%%%%%%%%%%%%%%%%%%%%
\section{Introduction}

Because of the limited computing power, current lattice computations of the hadronic matrix elements 
involving kaons are plagued by the necessity for introducing three important approximations: 
\begin{itemize}
\item[1.] The (partially) quenched approximation.
\item[2.] The extrapolation in the light quark masses: because of the inability to simulate directly with 
the physical $u/d$ quarks, one works with masses not lighter than about one half of that of the physical strange 
quark and then extrapolates to the physical $m_{u/d}$. Once the lighter quark masses are probed on the lattice, such 
as those announced in ref.~\cite{MILC}, the finiteness of the lattice volume becomes also an important approximation.
\item[3.] Degeneracy of valence quark masses in the kaon: matrix elements involving kaons are obtained 
with ``kaons" consisting of degenerate valence quarks whose mass is tuned in such a way as to produce 
a pseudoscalar meson with its mass equal to the physical $m_{K^0}=0.498$~GeV. 
\end{itemize}
In view of the great importance of the $\kkbar$ mixing amplitude in constraining the shape of the CKM unitarity 
triangle~\cite{CKM}, a quantitative estimate of the systematic errors induced by the above listed approximations 
is mandatory. 
That is where chiral perturbation theory (ChPT) enters the picture and offers a systematic approach for 
quantifying (at least roughly) the size of these errors. In ChPT one computes the coefficients of the  
chiral logarithms for various hadronic quantities in  order to (a) examine whether or not the quenched approximation 
introduces potentially large systematic error,  
(b) guide the chiral extrapolations, and (c) quantify the impact of 
the degeneracy in the quark masses on the evaluation of
the hadronic matrix elements. The coefficients of the 
chiral logarithms are predicted by quenched and full ChPT. Although   
convincing evidence for the presence of chiral logarithms in any numerical lattice data is still missing, 
a slight discrepancy from the linear (or quadratic) dependence on the variation 
of the light quark mass is occasionally observed. Before  
identifying such a discrepancy as an indication of the presence of chiral logarithmic behavior, 
one should make sure that the effects of finite volume are well under control. In particular, 
we would like to know how the finiteness of the lattice volume modifies the chiral logarithmic 
behavior of $f_K$ and $B_K$. In this paper we present the expressions obtained in 
three versions of ChPT that are relevant to present and future lattice simulations, i.e., in 
quenched ChPT (QChPT), partially quenched ChPT (PQChPT) and in full (standard) ChPT. 
Concerning the PQChPT, we will consider the case of $N_{\rm sea}=2$ degenerate 
dynamical quarks, which is the current practice in the lattice community. Those expressions, obtained 
in both the finite and infinite volumes, are then used to (i) show that chiral logarithmic 
behavior of $f_K$ and $B_K$ is indeed modified by the finiteness of the volume, and (ii) 
 assess the amount of systematic uncertainty induced by the finiteness of the lattice 
volume. As expected, finite volume effects increase as the mass of the 
valence light quark in the kaon, $m_q$, becomes smaller (we keep the strange quark mass fixed to its physical value). 
For quark masses $m_q \gtrsim m_s/3$ and volumes $V\gtrsim (2 \ {\rm fm})^3$, 
the finite volume effects are negligible. 
We will argue that even if one manages to push the quark masses 
closer to $m_{u/d}$, finite volume effects will start overwhelming the effects 
of the physical (infinite volume) chiral logs (unless one uses very large volumes). 
This unfortunately complicates the efforts currently being made in the lattice community to observe the 
chiral log behavior directly from the lattice data. Our finite volume ChPT formulae for $f_K$ and 
$B_K$ may be used to disentangle the finite volume effect from the physical chiral logarithmic 
dependence. Obviously this can be done only if the volume is sufficiently large and thus the finite volume 
corrections safely small enough to justify neglect of the unknown higher order corrections in the chiral expansion.

The remainder of this paper is organised as follows. In Sec.~\ref{sec2} we compute the chiral log corrections to 
$f_K$ and $B_K$ in all three versions of ChPT in infinite volume. In  Sec.~\ref{sec3} we discuss the same chiral 
corrections but in the finite volume. Both these sets of expressions are then combined in  Sec.~\ref{sec5} 
to examine the impact of the 
finite volume artefacts on the chiral behavior of $f_K$ and $B_K$. In Sec.~\ref{sec4} we discuss 
the finite volume effects on $f_K$ and $B_K$ and we briefly summarise in Sec.~\ref{sec6}.

\section{Results in infinite volume\label{sec2}}

To simplify the presentation and for an easier comparison with results available in the literature, we first 
briefly explain the notation adopted in this work and then present the expressions for the chiral logarithmic 
corrections that we computed in all three versions of ChPT.

\subsection{Chiral Lagrangians}

For the full (unquenched) ChPT we use the standard Lagrangian~\cite{GL,reviews}:
\begin{equation}
{\cal L}_{\rm ChPT}=\frac{f^2}{8}\textrm{tr}\,\bigl[(\partial_\mu\Sigma^\dagger)\,
(\partial^\mu\Sigma) + \Sigma^\dagger\chi + \chi^\dagger\Sigma\,\bigr] \,,
\end{equation}
with $f$ being the chiral limit of the pion decay constant, $f_\pi=132$~MeV. 
In addition,
\bea
 \chi &=& 2 \, B_0 \,{\cal M}= \frac{-2\la 0|\bar{u}u+\bar{d}d|0\ra}{f^{2}}\,{\cal M}\,,\nn \\ 
 {\cal M}&=&\textrm{diag}(m_u,m_d,m_s) \, , \nn \\
 \Sigma &=& {\mathrm{exp}}\left ( \frac{2i\Phi}{f}\right )\, , \\
 \Phi &=&  \left ( \begin{array}{ccc}
            \frac{\pi^{0}}{\sqrt{2}}+\frac{\eta}{\sqrt{6}} & \pi^{+} &  K^{+} \\
           \pi^{-} & -\frac{\pi^{0}}{\sqrt{2}}+\frac{\eta}{\sqrt{6}} & K^{0} \\
           K^{-} & \bar{K}^{0} & -\sqrt{\frac{2}{3}}\eta \end{array} \right ) \,. 
\eea
For the calculation in QChPT we will use the Lagrangian introduced in refs.~\cite{bg,sharpe92}:
\begin{equation}\label{qlagrangian}
{\cal L}_{\rm QChPT}=\frac{f^2}{8}\textrm{str}\,\left[(\partial_\mu\Sigma^\dagger)\,
(\partial^\mu\Sigma) + \Sigma^\dagger\chi + \chi^\dagger\Sigma\,\right] - m_0^2\Phi_0^2+\alpha_0\,(\partial_\mu\Phi_0)
(\partial^\mu\Phi_0)\ .
\end{equation} 
where $\Phi_0 \equiv \textrm{str}[\Phi]/\sqrt{6}$, is proportional to the graded extension of the
$\eta^\prime$, the trace over the chiral group indices has been replaced
by the supertrace over the indices of the graded group $SU(3|3)_L\times SU(3|3)_R$, and
the fields $\Sigma$ and $\chi$ are now graded extensions of $\Sigma$ and $\chi$, just defined above.

Finally, we choose the $SU(5|3)_L\times SU(5|3)_R$ setup for the PQChPT, 
i.e., three valence quarks ($u$, $d$, $s$), with masses 
$m_q\equiv m_{u}=m_{d}\neq m_{s}$,  and two degenerate sea quarks ($u_{\rm sea}$,\,$d_{\rm sea}$) 
of mass $m_{\rm sea}$. The Lagrangian is of the same form as the quenched one in eq.~(\ref{qlagrangian}), 
except that the indices now run over the graded group $SU(5|3)_L\times SU(5|3)_R$, and the fields 
$\Sigma$ and $\chi$ are extended to include the sea-quark sector~\cite{bernard-golterman}. Moreover, because of the presence 
of sea quarks, the $\eta^\prime$ decouples and $\Phi_0$ can be integrated out of the Lagrangian~\cite{shoresh}.

Throughout the paper, the evaluation of the chiral loop integrals is made by using 
na\" \i ve dimensional 
regularisation and the so called ``$\msbar +1$" renormalisation scheme of ref.~\cite{GL}.

\subsection{One-loop chiral log corrections to $f_K$ and $B_K$}

We begin by collecting the ChPT expressions for $f_K$ and $B_K$ in  infinite volume. 
We adopt the standard definition of the $B_K$ parameter, namely,
\bea\label{eq:def1}
B_K = {\langle \bar K^0\vert \bar s \gamma_\mu(1-\gamma_5)d\ \bar s \gamma_\mu(1-\gamma_5)d  \vert K^0 \rangle \over 
\displaystyle{\frac{8}{3}} \langle \bar K^0\vert \bar s \gamma_\mu(1-\gamma_5)d \vert 0\rangle \langle 0\vert \bar s \gamma_\mu(1-\gamma_5)d\vert K^0\rangle  }\,,
\eea
which is equal to $1$ in the vacuum saturation approximation.  The bosonised version of 
the relevant left-left ($\Delta S=2$) operator reads
\bea
 O_{27}^{\Delta S=2} = g_{27} {f^4\over 16} \left( \Sigma\partial_\mu \Sigma^\dagger\right)_{ds} 
 \left(\Sigma \partial^\mu \Sigma^\dagger \right)_{ds}\ .
\eea
To compute the chiral loop corrections to $f_K$, we use the standard bosonised left handed current: 
\bea
J_\mu^L= \bar s \gamma_\mu(1-\gamma_5)d \;\longrightarrow \;i {f^2\over 4} \left( \Sigma\partial_\mu \Sigma^\dagger\right)_{ds}\,.
\eea
In the following we will leave out the analytic terms (those accompanied by  
low energy constants) and focus only on the nonanalytic ones. As we will see, 
the analytic terms are not relevant to the discussion of finite volume effects.

The chiral logarithmic corrections to $f_K$ are
\bea\label{fkFULL}
\left({f_K\over f^{\rm tree}}\right)^{\rm ChPT}\!\!\!\!\!&=& 
1-\frac{3}{4 (4 \pi f)^2}\,\left[ m_\pi^2 \log \left(\frac{m_\pi^2}{\mu^2}\right) + 
2 m_K^2 \log (\frac{m_K^2}{\mu^2})+ 
m_\eta^2 \log \left(\frac{m_\eta^2}{\mu^2}\right) \right],\eea
\bea\label{fkpq}
\!\!\!\!\!\left({f_K\over f^{\rm tree}}\right)^{\rm PQChPT}\!\!\!\!\!&=&1-\frac{1}{2 (4 \pi f)^2}\,\left[ 
m_{SS}^2-m_K^2  + 
         \left( 2\,m_K^2 - m_{\pi}^2 + 
           m_{SS}^2 \right)  \log \left(\frac{m_{23}^2}{\mu^2}\right) \right. \nonumber \\
	  &&\hspace*{-6mm}\left. + {\left( m_{\pi}^2 + m_{SS}^2 \right) }\log 
	  \left(\frac{m_{13}^2}{\mu^2}\right)
	  - \frac{   
           m_K^2  m_{SS}^2  -m_\pi^2 \,\left( 2\,m_K^2 - m_\pi^2\right)
                }{2\,\left( m_K^2 - m_\pi^2 \right) }
	       \log \left(\frac{m_{22}^2}{m_\pi^2}\right)\, \right] ,\eea
\bea\label{fkq}
\!\!\!\!\!\left({f_K\over f^{\rm tree}}\right)^{\rm QChPT}\!\!\!=1-\frac{1}{3 (4 \pi f)^2}
\left[ \left( m_0^2 - \alpha_0 \,m_K^2 \right) -\frac{  m_0^2\,m_K^2 - \alpha_0 \,m_\pi^2\,
m_{22}^2 }{2\,
         \left( m_K^2 - m_\pi^2 \right) }\log \left(\frac{m_{22}^2}{m_\pi^2}\right) \right] ,
\eea
where ChPT, PQChPT, and QChPT stand for the full, partially quenched ($N_{\rm sea}=2$), and quenched 
chiral perturbation theory. In the above formulae, 
\bea\label{pokrate}
m_{SS}^2=2 B_0 m_{\rm sea}\ ,\quad && m_{22}^2 \equiv 2 B_0 m_s = 2 m_K^2 - m_\pi^2\ , \nn \\
m_{23}^2=B_0 (m_s+ m_{\rm sea}) ,&& m_{13}^2=B_0 (m_q+ m_{\rm sea})\,.
\eea 
We stress that we work in the exact isospin symmetry limit, i.e., $m_q\equiv m_u=m_d$. 
The results listed above agree with the ones available in the literature: eq.~(\ref{fkFULL}) was first obtained in 
ref.~\cite{GL}, eq.~(\ref{fkpq}) in refs.~\cite{Sharpe1997,leung1}, and eq.~(\ref{fkq}) in refs.~\cite{bg,sharpe92}.

For the $B_K$ parameter, we obtain:
\bea
\left({B_K\over B_K^{\rm tree}}\right)^{\rm ChPT}&=&
1  - \frac{2}{(4\pi f)^2}\,\left[ m_K^2 + m_K^2 \log \left(\frac{m_K^2}{\mu^2}\right) + 
       \frac{ m_\pi^2\,
          \left( m_K^2 + m_\pi^2 \right) }{4\,m_K^2}\log \left(\frac{m_\pi^2}{\mu^2}\right) \right.\nonumber \\ 
	  &&+\left.\frac{\left( 7\,m_K^2 - m_\pi^2 \right) \,
          m_\eta^2}{4\,m_K^2}\log \left(\frac{m_\eta^2}{\mu^2}\right) \right] \,,
\eea

\bea
\left({B_K\over B_K^{\rm tree}}\right)^{\rm PQChPT} \!\!\!\!\!\!\!\!&=& 1  - \frac{2}{(4\pi f)^2}\,\l\{
m_{SS}^2 + m_\pi^2 - \frac{m_K^4 + m_\pi^4}{2 m_K^2} 
+ m_K^2 
\l[ \log \left(\frac{m_K^2}{\mu^2}\right) + 2\log \left(\frac{m_{22}^2}{\mu^2}\right) \r]\r.\nn\\
&& \l. - \frac{1}{2}\l(
m_{SS}^2\frac{m_K^2 + m_\pi^2}{2 m_K^2} + 
m_\pi^2\frac{m_{SS}^2 - m_\pi^2}{ m_K^2 - m_\pi^2} \r)\log \left(\frac{m_{22}^2}{m_\pi^2}\right)
 \r\}, 
\eea

%\bea 
%\left({B_K\over B_K^{\rm tree}}\right)^{\rm QChPT}&=&1   - 
%\frac{1}{3 (4\pi f)^2}\,\left[ m_0^2 - \alpha_0 m_K^2   -
%\frac{m_0^2 m_K^2-\alpha_0 m_\pi^2m_{22}^2}{2 (m_K^2-m_\pi^2)}
%\log \left(\frac{m_{22}^2}{m_\pi^2}\right)\right].
%  \eea

\bea 
\left({B_K\over B_K^{\rm tree}}\right)^{\rm QChPT}&=&1  - 
  \frac{1}{3 (4 \pi f)^2}\,\left\{ 6 m_K^2 + 6 m_K^2\,\log \l(\frac{m_K^2}{\mu^2}\r) + 
       3 \ m_\pi^2\ \frac{m_K^2 + m_\pi^2}{m_K^2} \log \l(\frac{m_\pi^2}{\mu^2}\r) \r.\nn \\
      &&  \l.+ 3\ m_{22}^2\ \frac{m_{22}^2+m_K^2}{m_K^2}\, 
		\log \l(\frac{m_{22}^2}{\mu^2}\r)\,  \right. - {m_0^2} \,\left[ \frac{
           m_K^4 + \,m_{22}^2\,m_\pi^2}{ m_K^2 (m_K^2- m_\pi^2) }\,\log \l(\frac{m_{22}^2}{m_\pi^2}\r)\, - 4  \right] \nn \\
  && - {2 \alpha_0}\,\left[ {3  m_K^2} - \ \frac{m_{22}^2 \, m_\pi^2}{m_K^2}  + 
       \frac{m_\pi^2}{m_K^2}\ \frac{ m_K^4 + m_K^2\,m_\pi^2 - m_\pi^4}{ 
           m_K^2 - m_\pi^2  }\,\log \l(\frac{m_\pi^2}{\mu^2}\r)  \r.\nn\\
     &&\qquad \quad\l. \l.+\frac{m_{22}^2}{m_K^2}\ \frac{ m_K^4 + m_K^2\,m_{22}^2 - m_{22}^4  }{ 
            m_K^2 - m_{22}^2   }
			\,\log \l(\frac{m_{22}^2}{\mu^2}\r) \right]\r\}\,.
\eea
These results agree also with the ones previously computed in full ChPT~\cite{donoghue,bijnens-prades}, 
PQChPT~\cite{leung1}, 
and QChPT~\cite{sharpe92,leung2}, where more details about the actual calculation 
can be found.

\section{Results in finite volume\label{sec3}}

The calculation of the chiral logarithmic corrections in a finite box of volume $V=L^3$, with periodic 
boundary conditions,  is 
completely analogous to that in infinite volume, except for the fact that loop integrals 
now become sums over discretised three-momenta. As on the lattice, at the end of the calculation, 
the times of the kaon fields in the correlation function are sent to infinity. 
To abbreviate the expressions, we first introduce  
\bea\label{pokrate2}
\omega_\pi^2&=&{\vec q\ }^2+m_\pi^2\,, \qquad \omega_K^2={\vec q\ }^2+m_K^2\,, 
\qquad \omega_{22}^2={\vec q\ }^2+m_{22}^2 \,, 
\eea
and analogously for $\omega_{13}$, $\omega_{23}$, with the corresponding masses already defined in 
eq.~(\ref{pokrate}).
As in infinite volume, $m_0$ and $\alpha_0$ are the $\eta^\prime$ parameters of the quenched 
theory.

For the decay constant $f_K$ in all three versions of the ChPT, we obtain,  
\bea\label{fkFULLFV}
\!\!\!\!\!\left({f_K\over f^{\rm tree}}\right)^{\rm ChPT}\!\!\!\!\!&=& 
1-\frac{3}{8 f^2 L^3}\,\sum_{\vec q} \left( {1 \over \omega_\pi}\ + \ {2 \over \omega_K}\ +\ {1 \over \omega_\eta}\right), \\
&&\hfill \nonumber \\  
&&\hfill \nonumber \\
\!\!\!\!\!\left({f_K\over f^{\rm tree}}\right)^{\rm PQChPT}\!\!\!\!\!\!&=&1 + 
\frac{1}{8 f^2 L^3}\sum_{\vec q}\ 
\left[ \frac{ m_{SS}^2 - m_\pi^2}{2\ \omega_\pi^3} +\frac{ m_{SS}^2 - m_{22}^2}{2\ \omega_{22}^3} - 
      4  \left( {1\over \omega_{13}} + {1\over \omega_{23}} \right)
      \r.\nn\\ 
&&\qquad\qquad\qquad\qquad\l.      +  \frac{m_{SS}^2 - m_K^2}{m_K^2 - m_\pi^2 } 
\left( {1\over \omega_{22}} - {1\over \omega_\pi} \right)  \right]  , \\
&&\hfill \nonumber \\
&&\hfill \nonumber \\ 
\label{fkqFV}
\!\!\!\!\!\left({f_K\over f^{\rm tree}}\right)^{\rm QChPT}\!\!\!\!\!&=&1 - \frac{1}{24 f^2 L^3}\ 
\sum_{\vec q}\ 
     \left\{ m_0^2\ \left[
     \frac{ 2}{ m_K^2 - m_\pi^2 } \left( \frac{1}{ \omega_\pi} -\frac{1}{ \omega_{22}} \right)  -\frac{1}{ \omega_{22}^3}- 
          \frac{1}{\omega_\pi^3} \right] \right. \nonumber \\ 
	  &&\hspace*{27mm}\left.- \alpha_0 \ \left[
     \frac{ 2 m_K^2}{ m_K^2 - m_\pi^2 } \left( \frac{1}{\omega_\pi} -\frac{1}{\omega_{22}} \right)  -
     \frac{m_{22}^2}{ \omega_{22}^3}- 
          \frac{m_\pi^2}{\omega_\pi^3} \right] \right\}  ,
\eea
while for the $B_K$ parameter we have
\beq
\left({B_K\over B_K^{\rm tree}}\right)^{\rm ChPT}=1+\frac{1}{4 f^2 L^3}{\sum_{\vec q}}\,\left[ 
 \frac{2 m_K^2}{ \omega_K^3} \ -\ \frac{m_K^2 + m_\pi^2}{m_K^2\ \omega_\pi} \ -\ \frac{7 m_K^2 - m_\pi^2}{m_K^2\ \omega_\eta}  \right] \,,
\eeq
\bea
\left({B_K\over B_K^{\rm tree}}\right)^{\rm PQChPT}&=&1+\frac{1}{4 f^2 L^3}{\sum_{\vec q}}\,\left[ 
\frac{\left(m_K^2 + m_\pi^2 \right)\ 
      \left( m_{SS}^2 - m_\pi^2 \right) }{2\ m_K^2\ \omega_\pi^3}
+\frac{\left(m_K^2 + m_{22}^2 \right)\ 
      \left( m_{SS}^2 - m_{22}^2 \right) }{2\ m_K^2\ \omega_{22}^3}\r. \nn \\
&& \qquad \qquad  \qquad+\frac{ 2 m_K^2}{\omega_K^3}
-\frac{ m_K^4 - m_\pi^4 +  2  m_K^2 (m_{SS}^2 - m_\pi^2 )}{m_K^2 \omega_{\pi}(m_K^2-m_\pi^2)}\nn\\
&& \qquad \qquad  \qquad \l.  + 
\frac{ m_K^4 - m_{22}^4 +  2  m_K^2 (m_{SS}^2 - m_{22}^2 )}{m_K^2 \omega_{22}(m_K^2-m_\pi^2)} \right]\,,
\eea
\bea
\left({B_K\over B_K^{\rm tree}}\right)^{\rm QChPT}&=&1- \frac{1}{2 f^2 L^3}{\sum_{\vec q}}\,\left\{
\frac{ m_\pi^2 +  m_K^2}{m_K^2}{1\over \omega_\pi} + 
\frac{ m_{22}^2 +  m_K^2}{m_K^2}{1\over \omega_{22}} -  \frac{ m_K^2}{\omega_K^3}  \r. \nn \\
	  &&\quad -  {m_0^2\over 6} 
       \left[ \frac{m_{22}^2 + m_K^2}{ m_K^2\ \omega_{22}^3} + 
         \frac{m_\pi^2 + m_K^2}{ m_K^2\ \omega_\pi^3} - 
         \frac{4}{  m_K^2 - m_\pi^2 }\left( {1\over \omega_\pi} - {1\over \omega_{22}}  \right) \right]  \nn \\
      &&\quad + {\alpha_0 \over 6 m_K^2} \ \left[ 
      \frac{m_{22}^2 ( m_{22}^2 + m_K^2)}{\omega_{22}^3} +
      \frac{m_{\pi}^2 ( m_{\pi}^2 + m_K^2)}{\omega_{\pi}^3}\r. \nn \\ 
      && \l.\l.\qquad\qquad\quad - 
      {2 (m_K^4 + m_\pi^2 m_{22}^2)\over m_K^2-m_\pi^2} 
      \left( {1\over \omega_\pi} - {1\over \omega_{22}} \right) \right]
  \right\} \,.
\eea
We are now faced with the problem of evaluating the sums over discrete momenta $\vec q = \frac{2\pi}{L}\vec n$, with $\vec n \in
Z^3$.

\subsection{Evaluation of the chiral loop sums\label{ingredients}}

The sums that appear in the calculation of the tadpole diagrams are of the form
\beq
\frac{1}{L^3}\sum_{\vec q} \frac{1}{({\vec q\ }^2+M^2)^s}\,,
\eeq
where $M$ stands for the generic mass. It is very easy to verify that
\beq
\lim_{L\to\infty}\frac{1}{L^3}\sum_{\vec q} \frac{1}{({\vec q\ }^2+M^2)^s} =
	\frac{\sqrt{4 \pi}\ \Gamma(s+\frac12)}{\Gamma(s)} \int \frac{d^4 q}{(2\pi)^4} \frac{1}{(q^2+M^2)^{s+\frac12}} \,.
\eeq
For finite $L$, one can write
\beq
\frac{1}{L^3}\sum_{\vec q} \frac{1}{({\vec q\ }^2+M^2)^s}= 
\frac{\sqrt{4 \pi}\  \Gamma(s+\frac12)}{\Gamma(s)} \int \frac{d^4 q}{(2\pi)^4} \frac{1}{(q^2+M^2)^{s+\frac12}}+
\xi_s(L,M) \,, \label{eq:defxi}
\eeq 
where $\xi_s(L,M)$ is simply the difference between the finite volume sum and the infinite volume integral. 
This function is finite and needs no regularization since it represents an infrared effect. In other words,  
the integral and the sum diverge in the same way. Equation~\ref{eq:defxi} can then be considered as a way 
to regularize the sums which, in addition, allows us to adopt the same renormalization scheme for both  
integrals and sums.

In the following few steps, we show how $\xi_s(L,M)$ is simplified to
\bea
\xi_s(L,M)&=&
\frac{1}{L^3}\sum_{\vec q} \frac{1}{({\vec q\ }^2+M^2)^s}-
\frac{\sqrt{4 \pi}\  \Gamma(s+\frac12)}{\Gamma(s)} \int \frac{d^4 q}{(2\pi)^4} \frac{1}{(q^2+M^2)^{s+\frac12}}\cr
&=& \frac{1}{\Gamma(s)}
\int_0^\infty d\tau\, \tau^{s-1} e^{-\tau M^2} \frac{1}{L^3}\sum_{\vec q} e^{-\tau {\vec q\ }^2} -
\frac{1}{\Gamma(s)} \int_0^\infty d\tau \tau^{s-1} e^{-\tau M^2}\int {d^3q\over (2\pi)^3} e^{-\tau {\vec q\ }^2}\cr
&=&\frac{1}{\Gamma(s)}\int_0^{\infty}d\tau\, \tau^{s-1}e^{-\tau M^2} 
\l\{\l[\frac{1}{L}\vartheta\l(\frac{4\pi^2\tau}{L^2}\r)\r]^3-\frac{1}{8(\pi \tau)^{3/2}} \r\}\label{e1} \\
&=&\frac{L^{2s-3}}{(2\pi)^{2s}\Gamma(s)}\int_0^{\infty} d\tau\, \tau^{s-1} e^{-\tau \l(\frac{ML}{2\pi}\r)^2}
\left\{\left[\vartheta(\tau)\right]^3-\l(\frac{\pi}{\tau}\r)^{3/2}\right\} \label{e1b}\nn ,
\eea
where the elliptic theta function $\vartheta(\tau)$ is defined as~\footnote{
The function $\vartheta(\tau)$ is obtained from the commonly used function $\vartheta_3(u,q) = \displaystyle{\sum_{n=-\infty}^{\infty}}q^{n^2} e^{2nui}$, after
replacing, $u=0$ and $q= e^{-\tau}$. For the numerical analysis, we use the function 
predefined in {\sc Mathematica}, namely,   
{\tt  EllipticTheta[3, 0, $\tt e^{-\tau}$]}. For more details on the elliptic functions, see ref.~\cite{grad}. 
}
\bea
\vartheta(\tau)&\equiv& \sum_{n=-\infty}^{\infty} e^{-\tau\, n^2}
\eea
and satisfies the Poisson summation formula~\cite{poisson}
\bea
\vartheta(\tau) =\sqrt{\pi\over \tau}\ \vartheta\left({\pi^2\over \tau}\right)\ .\label{eq:poisson}
\eea

Applying the formula~(\ref{eq:poisson}) to eq.~(\ref{e1}), we get
\bea
\xi_s(L,M)&=&
\frac{1}{(4\pi)^{3/2}\Gamma(s)}\int_0^{\infty}d\tau\ \tau^{s-5/2}e^{-\tau M^2} 
\l[  \vartheta^3\l(\frac{L^2}{4\tau}\r)-1  \r]. 
\eea
In the asymptotic limit $L\to \infty$, the theta function behaves as 
$\vartheta(L^2/4\tau ) \sim 1 + 2 e^{-L^2/4\tau}$, so that in the same limit we can write
\bea\label{eq:LL}
\xi_s(L,M) \to \frac{3\sqrt{\pi}}{\Gamma(s)(2\pi)^{3/2}}\frac{e^{-ML}}{(ML)^{2-s}}(2M^2)^{3/2-s}.  
\eea

\section{Impact of finite volume effects on the chiral behavior of $f_K$ and $B_K$\label{sec5}}

In recent years considerable effort has been invested in controlling the chiral extrapolations 
of the hadronic matrix elements computed on a lattice. To guide the extrapolation from the 
directly accessible  quark masses, $r\approx 0.5$, down to the physical $r\to r_{u/d}=0.04$, 
one can rely on the expressions obtained in ChPT (quenched, partially quenched, or full). 
Those expressions, however, contain chiral logarithmic terms which so far have not been 
observed in the numerical studies. 
An important task before the lattice community is to lower the quark mass and get closer 
to the region in which the chiral logarithms become clearly visible.
\begin{figure}
\begin{center}
\hspace*{-3mm}\epsfig{file=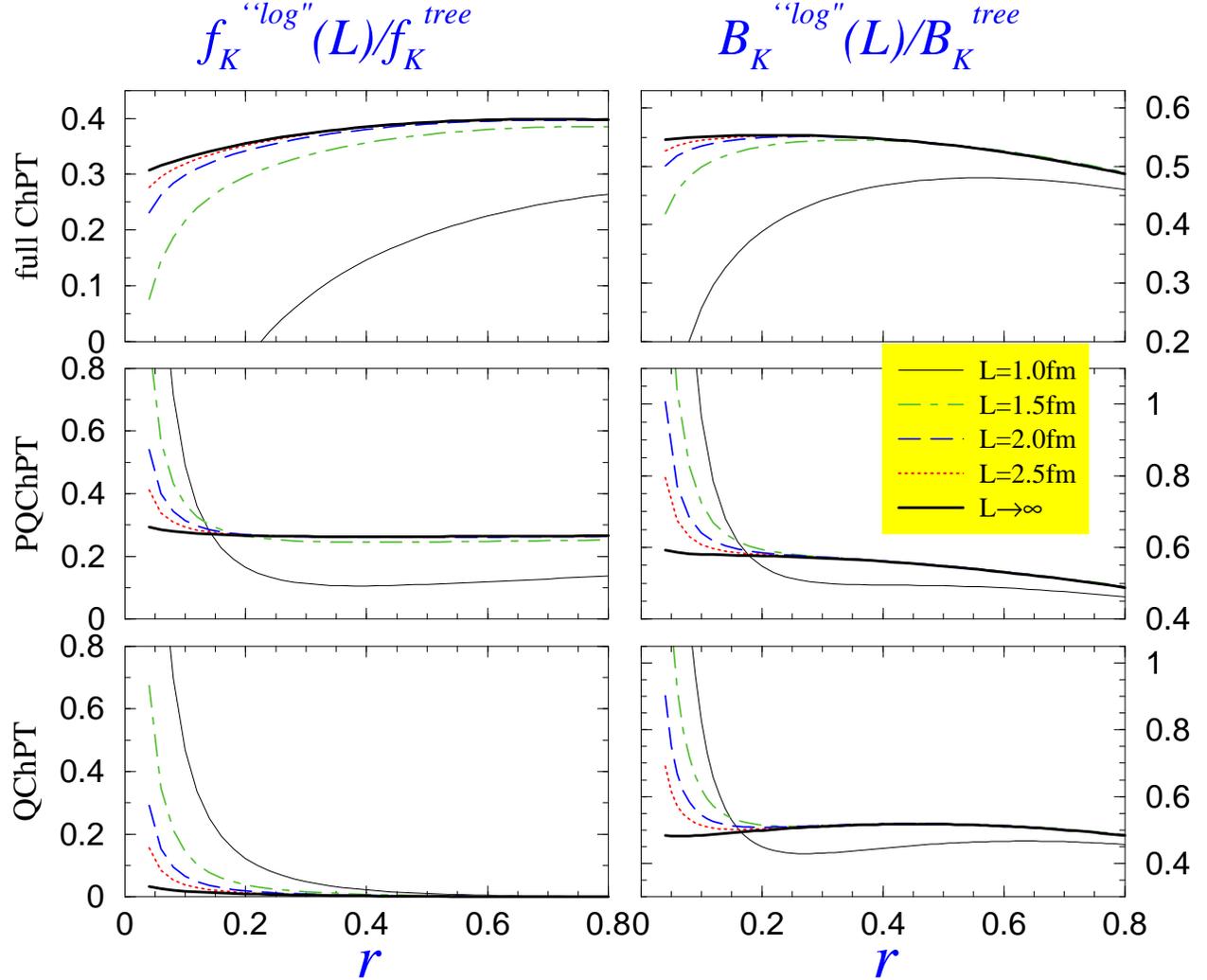, height=14.6cm}
\caption{\label{fig4}\footnotesize{\sl 
From top to bottom, we plot the chiral logarithmic corrections as predicted in full, 
partially quenched ($r_{\rm sea}=m_{\rm sea}/m_s^{\rm phys}=0.5$) and quenched ChPT, respectively, as  
functions of the 
light valence quark mass $r=m_q/m_s$, where the strange quark 
mass is fixed to its physical value. In each plot the thick line corresponds to 
the physical (infinite volume) chiral logarithm, whereas the other four curves correspond 
to the logarthmic contributions computed in the finite volume
$V=L^3$, where for $L$ we choose the values shown in the legend. The renormalisation scale is chosen to be
$\mu=1$~GeV.}
} 
\end{center}
\end{figure}
However, by decreasing the quark mass, the sensitivity to the finiteness of the lattice box of side 
$L$ becomes more pronounced. Moreover, the finite volume effects  modify the nonlinear 
light quark dependence in the same way, i.e., they enhance the chiral logs. The problem is that 
the nonlinearity induced by the finite volume is larger than that due to the presence 
of physical chiral logarithms. To illustrate that statement, in fig.~\ref{fig4} we plot the chiral 
log contributions in the finite and infinite volumes, by using the expressions presented in the 
previous section for both $f_K$ and $B_K$ in all three versions of ChPT. From that plot we see that 
it is very difficult to distinguish between physical chiral logarithms (thick curves) 
and the finite volume effect, even if one manages to work with very light quarks
on the currently used lattice volumes.  
For smaller masses, at which the chiral logarithms are expected to set in, the finite volume effects 
completely overwhelm the physical non-linearity. 

A possible way out would be to fit the lattice data to the finite volume forms (see Sec.~\ref{sec3}) and not 
to those of the infinite volume, given in Sec.~\ref{sec2}. That, of course, is legitimate if one 
assumes the validity of the NLO ChPT formulae. Finally, the curves corresponding to $L=1$~fm should be taken 
cautiously because this volume may be too small for ChPT to set in, as recently discussed in ref.~\cite{durr}.

\section{Finite volume corrections \label{sec4}}

In this section we combine the formulae derived in Secs.~\ref{sec2} and \ref{sec3} to 
discuss the shift of $f_K$ and $B_K$ induced by the finite volume effects. 
Before embarking on this issue, we first briefly remind the reader about the similar shift 
in the case of the pion mass where, for large $L$,  the one-loop ChPT expression indeed agrees 
with  the general formula derived by L\"uscher in ref.~\cite{luscher}. 
Since the analogous general formulae for $f_K$ and $B_K$ do not exist, we will derive 
them by taking the large $L$ limit of our one-loop ChPT formulae.

\subsection{Contact with L\"uscher's formula}

To make contact with L\"uscher's formula, we subtract the one-loop chiral correction to the 
pion mass (squared) as obtained in the full (unquenched) ChPT in infinite volume from the one 
obtained in a finite volume. By evaluating the sums, as described in the previous section, we have
\bea\label{luscherChPT}
&&m_\pi^2(\infty) = 2 B_0 m_q \left\{ 1 + {1 \over (4\pi f)^2}\left[ m_\pi^2 \log\left({m_\pi^2\over \mu^2}\right)
- {1\over 3} m_\eta^2 \log\left({m_\eta^2\over \mu^2}\right)\right] \right\} \,, \nonumber \\
&&m_\pi^2(L) = 2 B_0 m_q \left[ 1 + {1 \over 2 f^2 L^3}\sum_{\vec q} \left( 
{1\over \omega_\pi}\ -\ {1\over 3\ \omega_\eta}\right) \right] \,, \nonumber \\
\Longrightarrow&& {\Delta m_\pi^2\over m_\pi^2} \equiv {m_\pi^2(L) - m_\pi^2(\infty) \over m_\pi^2(\infty)}= {1 \over 2 f^2} \left[ \xi_{1/2}(L,m_\pi)-{1\over 3} \xi_{1/2}(L,m_\eta) \right]\,,
\eea
which coincides with the result of ref.~\cite{GL2}.~\footnote{Notice that the function 
$g_r(M^2,0,L)$, defined in ref.~\cite{GL2}, is related to $\xi_{s}(L,M)$ through
\beq
\xi_s(L,M)\ =\ { \sqrt{4 \pi} \over \Gamma(s)} \ g_{s+1/2}(M^2,0,L)\,.\nn
\eeq} Analytic terms in  $m_\pi^2(\infty)$ and $m_\pi^2(L)$ were omitted since they cancel in 
$(\Delta m_\pi^2)/m_\pi^2$.

To recover L\"uscher's formula, one takes the limit $L\to \infty$, which amounts to 
using the asymptotic form~(\ref{eq:LL}) in eq.~(\ref{luscherChPT}),  
\bea\label{luscher}
{\Delta m_\pi\over m_\pi} \simeq {3\over 2}\left({m_\pi\over f}\right)^2 {e^{-m_\pi L} \over (2 \pi m_\pi L)^{3/2}} 
\,,
\eea
where only the leading exponential has been kept.
The benefit of eq.~(\ref{luscherChPT}) is that it offers insight into the subleading terms, suppressed 
by higher powers in $e^{-m_\pi L}$ in L\"uscher's formula. In the range of volumes in which  the right-hand 
side of eq.~(\ref{luscher}) becomes sizable, the subleading exponential terms cannot be neglected and 
the formula~(\ref{luscherChPT}) has to be used.  For the volumes currently used in lattice simulations, 
these corrections are important if one is to work with very light pions.

Before closing this subsection, two important comments are in order, though. 
First, L\"uscher's formula relates the finite volume 
shift of the pion mass to the $\pi\!-\!\pi$ scattering amplitude. Equation~\ref{luscher} refers 
to the tree level $\pi\!-\!\pi$ scattering amplitude. A recent study in ref.~\cite{durr} 
shows that the inclusion of the NLO chiral corrections to the $\pi\!-\!\pi$ scattering amplitude 
produces a sizable correction to eq.~(\ref{luscher}). It is, however, not clear whether or not such 
a conclusion persists in the full (nonasymptotic) case, i.e., eq.~(\ref{luscherChPT}). It is even less 
clear if such a conclusion carries over to other quantities. To resolve that issue, one should 
compute the finite volume two-loop chiral correction to the pion mass (and to other quantities), 
which is beyond the scope of the present work. 
It is clear, however, that before this point is clarified, one cannot safely use the one-loop
calculation to correct for the finite volume effects. At present the one-loop ChPT finite volume 
expressions are useful for making a rough estimate of the finite volume corrections.  
The second comment is that the derivation of L\"uscher's formula relies crucially on unitarity. 
Since the unitarity in the partially quenched and quenched theories is lost,
L\"uscher's formula is not valid in these theories.

\subsection{Finite volume corrections to $f_K$}

We now use the expressions for the decay constant $f_K$, derived in the infinite [eqs.~(\ref{fkFULL})-(\ref{fkq})] 
and finite [eqs.~(\ref{fkFULLFV})-(\ref{fkqFV})] volume cases,     
to estimate the shift in $f_K$ due to the finiteness of the volume. For that purpose 
we define 
\beq\label{ratiofk}
{\Delta f_K\over f_K}\equiv{f_K(L) - f_K(\infty)\over f_K(\infty)}\,.
\eeq
It should be clear that the analytic terms multiplied by the low energy constants 
(omitted in Secs.~\ref{sec2} and~\ref{sec3}) cancel in the 
ratio~(\ref{ratiofk}).~\footnote{
Written schematically, $f_K(\infty)= f^{\rm tree}(1+ {\log_{\infty}} + C m_q)$ 
is equivalent to 
$f^{\rm tree}= f_K(\infty) (1-  \log_{\infty}  - C m_q)$, where the analytic term is multiplied by the 
 generic low energy constant $C$. $f^{\rm tree}$ is the same in finite and infinite 
volumes, so that one simply obtains $f_K(L)=f_K(\infty) (1-  {\log_{\infty}} +  {\log_{L}})$. 
Thus, at this order, the effects of low energy constants cancel.}
Finally, we have
\beq\label{fK1}
\left({\Delta f_K\over f_K}\right)^{\rm ChPT}=
-\frac{3}{8\,f^2}\,\left[  
      {{\xi }_{\frac{1}{2}}}(L,{m_{\pi }}) \ +\ {2}\ {{\xi }_{\frac{1}{2}}}(L,{m_K})\ + \ 
      {{\xi }_{\frac{1}{2}}}(L,{m_{\eta }}) \right] \,,
\eeq

\bea\label{fK2}
\left({\Delta f_K\over f_K}\right)^{\rm PQChPT}&=& \frac{1}{8\,f^2}\,\left\{ 
\frac{  m_{SS}^2  - m_\pi^2 }{ 2 } {{\xi }_{\frac{3}{2}}}(L,{m_{\pi }}) 
+ \frac{  m_{SS}^2  - m_{22}^2 }{ 2 } {{\xi }_{\frac{3}{2}}}(L,{m_{22}}) - 4\ \l[ {{\xi }_{\frac{1}{2}}}(L,{m_{13}}) \r.\r.\nn\\
&& \qquad \quad\l. \l. + 
       {{\xi }_{\frac{1}{2}}}(L,{m_{23}})\r] 
+ \frac{  m_{SS}^2  - m_K^2 }{ m_K^2 - m_\pi^2 }\l[{{\xi }_{\frac{1}{2}}}(L,{m_{22}}) 
- {{\xi }_{\frac{1}{2}}}(L,{m_{\pi}}) \r] \r\} 
 \,,
\eea

\bea\label{fK3}
\left({\Delta f_K\over f_K}\right)^{\rm QChPT}&=&-\frac{1}{12\,f^2}\left\{ \frac{ 
 \alpha_0 \,m_K^2 -m_0^2 }{m_K^2-m_\pi^2 }\,
        \left[ {{\xi }_{\frac{1}{2}}}(L,{m_{22}}) - {{\xi }_{\frac{1}{2}}}(L,{m_{\pi }})\right] 
\r.\nn \\ && \l.
       - \frac{  m_0^2 -\alpha_0 \,m_{22}^2   }{2}\,{{\xi }_{\frac{3}{2}}}(L,{m_{22}}) 
	   - \frac{ m_0^2 - \alpha_0 \,m_\pi^2  }{2} \,
         {{\xi }_{\frac{3}{2}}}(L,{m_{\pi }})\right\}\,. 
\eea
Similar expressions in full and quenched ChPT, but for the case in which the kaon consists of two  
quarks degenerate in mass,  were obtained in ref.~\cite{sharpe92}. 
For the general nondegenerate case and for PQChPT, the above formulae are new. In ref.~\cite{aubin},  
the finite volume terms are taken into account while computing the full ChPT corrections to $f_K$ relevant to the 
lattice computation of this quantity with staggered quarks.

\begin{figure}[t!]
\begin{center}
\epsfig{file=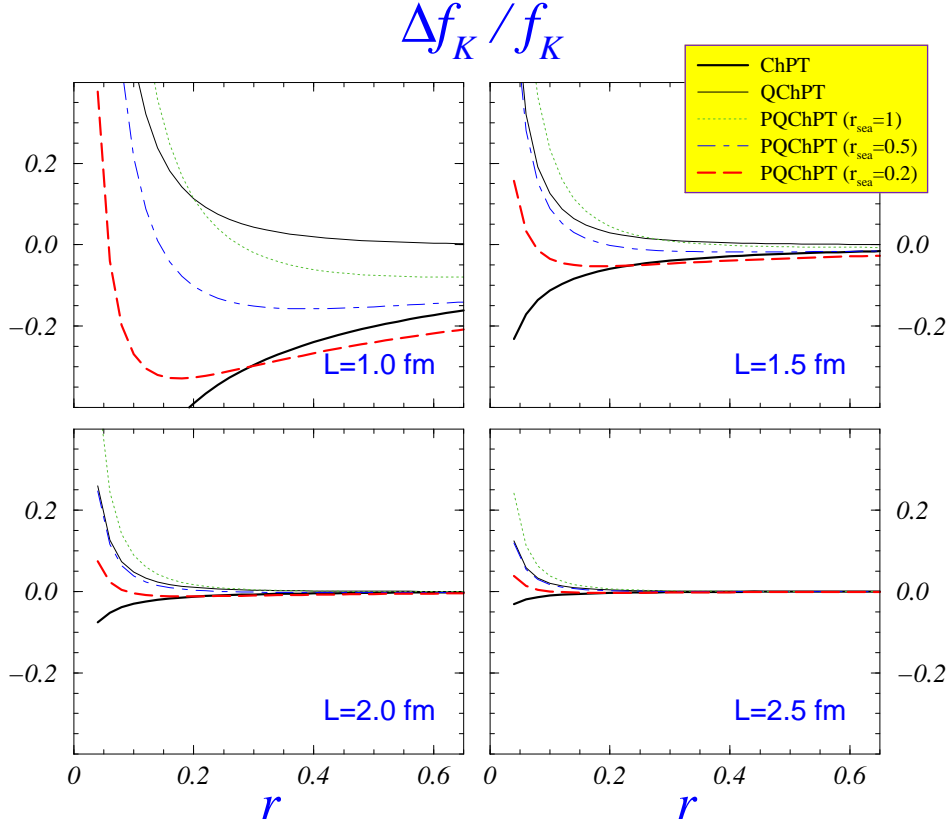, height=11.cm}
\caption{\label{fig1}\footnotesize{\sl 
The finite volume corrections to $f_K$ in full, partially quenched and quenched theory, 
eqs.~(\ref{fK1}),(\ref{fK2}),(\ref{fK3}), respectively. The partially quenched case that we consider 
is the one with $N_{\rm f}=2$ dynamical quarks degenerate in mass for which we take $r_{\rm
sea}=m_{\rm sea}/m_s^{\rm phys} = 1, 0.5, 0.2$. Each plot corresponds to a different value of the size of the 
side of the box $L$, indicated in the plots. We keep the same scale, to better appreciate the reduction 
of the finite volume effects as $L$ is increased.}} 
\end{center}
\end{figure}

In fig.~\ref{fig1}, we illustrate the finite volume effects as predicted by the above formulae. 
Since the mass of the strange quark is directly simulated on the lattice, we keep it fixed.  
As for the light quark we 
define, $r=m_q/m_s^{\rm phys}$, which we then  vary as $r\in [r_{u/d}, 1]$, where 
$r_{u/d}=(m_u+m_d)/2m_s = 0.04$~\cite{leutwyler}. We also use the Gell-Mann--Oakes--Renner (GMOR) and 
Gell-Mann--Okubo formulae, namely, 
\bea
m_\pi^2 = 2 B_0 m_s r\ ,\quad m_K^2 = 2 B_0 m_s {r+1\over 2}\ , \quad m_\eta^2 = 2 B_0 m_s {r+2\over 3}\ . 
\eea
The illustration in fig.~\ref{fig1} is made for realistic volumes   
currently used in lattice simulations: $L \in [1\ {\rm fm},\ 2.5\ {\rm fm}]$. 
In obtaining the quenched curves, we assume $m_0=0.6$~GeV and  $\alpha_0=0.05$. The curves are insensitive to 
the value of $\alpha_0$. On the other hand, the effect shown in fig.~\ref{fig1} becomes pronounced 
if $m_0$ is increased to $m_0=0.65$~GeV or $m_0=0.7$~GeV, values sometimes also quoted in the literature.

From fig.~\ref{fig1} we see that the finite volume effects become more pronounced as the light quark gets 
closer to the physical $u/d$ quark mass. In particular, they result in shifting the quenched $f_K$ toward a larger value, 
wheras the shift of $f_K$ in full (unquenched) QCD is opposite, i.e., the finite volume effects lower the value of $f_K$. 
 The partially unquenched cases lie between the two. We see that 
the quenched chiral logs become dominant as soon as the mass of the valence quark 
becomes lighter than the sea-quark mass. We did not plot the case when the valence and 
 sea quarks are degenerate since such a curve is very close 
to the full ChPT case. It is worth noticing that in the region of $r\lesssim 0.2$ the finite volume 
effects for the partially quenched $f_K$, with $r_{\rm sea}=1$, are larger than for the quenched case.

\subsection{Finite volume corrections to $B_K$}

We proceed in a completely analogous way as in the last subsection and define
\beq
{\Delta B_K\over B_K}\equiv{B_K(L) - B_K(\infty)\over B_K(\infty)}\,.
\eeq
The corresponding ChPT expressions read
\bea\label{BK1}
\left({\Delta B_K\over B_K}\right)^{\rm ChPT}
&=&\frac{1}{4 f^2} \l[- \frac{m_K^2+m_\pi^2}{m_K^2} \ {{\xi }_{1/2}}(L,{m_{\pi }}) + 
    2 m_K^2\ {{\xi }_{3/2}}(L,{m_K})\r.\nn \\
    &&\qquad\, \l.   - 
    \left( 7- \frac{m_\pi^2}{m_K^2} \right) \ {{\xi }_{1/2}}(L,{m_{\eta }})  \r] \,, 
\eea
\bea\label{BK2}
\left({\Delta B_K\over B_K}\right)^{\rm PQChPT}
\!\!\!\!&=& - \frac{1}{2 f^2}\l[
\left( {5 m_K^2-m_\pi^2\over 2 m_K^2}\ -\ { m_{SS}^2 - m_K^2\over m_K^2-m_\pi^2}\right)\ {{\xi }_{1/2}}(L,{m_{22}}) \r. \nn \\ 
	   &&\quad + 
	   \left( { m_K^2+m_\pi^2\over 2 m_K^2}\ +\ {m_{SS}^2 - m_\pi^2\over m_K^2-m_\pi^2}\right)\ {{\xi }_{1/2}}(L,{m_{\pi }}) \nn \\
	   &&\quad +\, \frac{\left( m_K^2 + m_{22}^2 \right) \,\left( m_{22}^2 - m_{SS}^2 \right)}
	   	{4 m_K^2}\,{{\xi }_{3/2}}(L,{m_{22}}) \nn \\
	   &&\quad \l. - m_K^2\ {{\xi }_{3/2}}(L,{m_K})
	   -\, \frac{\left( m_K^2 + m_\pi^2 \right) \,\left( m_{SS}^2 - m_\pi^2 \right)}{4 m_K^2}\,
       {{\xi }_{3/2}}(L,{m_{\pi }})\r]\,,
\eea

\bea\label{BK3}
\left({\Delta B_K\over B_K}\right)^{\rm QChPT}
\!\!\!\!\!\!&=&\frac{1}{2 f^2}\l\{
\l[{2 m_0^2\over 3(m_K^2-m_\pi^2)} - {3 m_K^2 - m_\pi^2\over m_K^2} - {\alpha_0\over 3}\left( 
{m_K^2+m_\pi^2\over m_K^2-m_\pi^2} + {m_\pi^2\over m_K^2}\right)\r]\ {{\xi }_{1/2}}(L,{m_{22}}) \r. \nn \\
	&&\quad 
- \l[{2 m_0^2\over 3(m_K^2-m_\pi^2)} + {m_K^2 + m_\pi^2\over m_K^2} - {\alpha_0\over 3}\left( 
{m_K^2+m_\pi^2\over m_K^2-m_\pi^2} + {m_\pi^2\over m_K^2}\right)\r]\ {{\xi }_{1/2}}(L,{m_{\pi }}) \nn \\
	&&\quad +\frac{\left( m_K^2 + m_{22}^2 \right) \,
       \left( m_0^2 - \alpha_0 \ m_{22}^2  \right) }{6\ 
       m_K^2}\,{{\xi }_{3/2}}(L,{m_{22}})  +  {m_K^2}\ {{\xi }_{3/2}}(L,{m_K}) \nn \\
	&&\quad \l. + 
    \frac{\left( m_K^2 + m_\pi^2 \right) \,\left( m_0^2 - \alpha_0 \ m_\pi^2 \right)}
	{6\ m_K^2}\ {{\xi }_{3/2}}(L,{m_{\pi }})\r\}\,.
\eea
\begin{figure}[t!]
\begin{center}
\epsfig{file=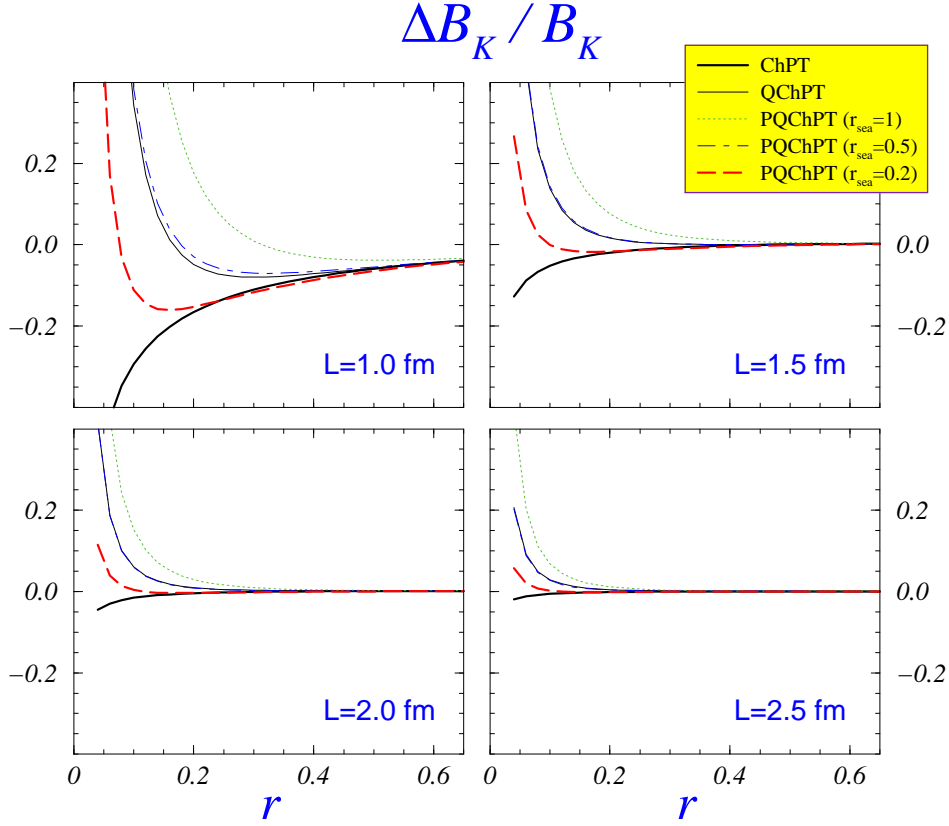, height=11.cm}
\caption{\label{fig2}\footnotesize{\sl 
The finite volume corrections to $B_K$ in full, partially quenched and quenched theory, 
eqs.~(\ref{BK1}),(\ref{BK2}),(\ref{BK3}), respectively. } }
\end{center}
\end{figure}
The illustration, similar to the one discussed in the previous subsection, is provided in fig.~\ref{fig2}. 
We observe that, as in the case of $f_K$, the finite volume effects become pronounced as 
the light valence quark is lowered toward the physical $u/d$ quark mass. Moreover, they 
have a tendency to enhance the nonlinearities which would otherwise be attributed to the 
physical chiral logarithms.~\footnote{By ``physical chiral logarithms", we mean 
the chiral logarithmic behavior in infinite volume.}

\subsection{Asymptotic $L\to \infty$ formulae for $f_K$ and $B_K$}

As discussed at the beginning of this section, the difference of the one-loop ChPT formulae for the pion 
mass in the finite and infinite volume (i) reproduces L\"uscher's general formula at LO, and 
(ii) allows one to estimate the size of the corrections suppressed in L\"uscher's formula.
Similar asymptotic formulae for $f_K$ and $B_K$ can be obtained from our full ChPT expressions for  
$\Delta f_K/f_K$ and $\Delta B_K/B_K$, i.e., eqs.~(\ref{fK1}) and~(\ref{BK1}),
respectively. By using the asymptotic form~(\ref{eq:LL}), we get
\bea\label{asymptotic}
{\Delta f_K\over f_K} \simeq -{9\over 4}\left({m_\pi\over f}\right)^2 {e^{-m_\pi L} \over (2 \pi m_\pi L)^{3/2}}\,,\quad
{\Delta B_K\over B_K} \simeq -{3\over 2}\frac{m_K^2+m_\pi^2}{m_K^2} \left({m_\pi\over f}\right)^2 {e^{-m_\pi L} \over (2 \pi m_\pi L)^{3/2}}
  \,.
\eea
To check how good an approximation these formulae are to the complete ones, 
given in eqs.~(\ref{fK1}) and~(\ref{BK1}), we made a numerical comparison of the two, and 
conclude that for volumes larger than $(2\ {\rm fm})^3$ and masses $r \gtrsim 1/4$, 
eq.~(\ref{asymptotic}) is an excellent approximation. Otherwise, i.e., in the region in which the  
finite volume effects become important, the asymptotic forms~(\ref{asymptotic}) become 
inadequate and eqs.~(\ref{fK1}) and~(\ref{BK1}) should be used. We note, in
passing, that formulae similar to those in eq.~(\ref{asymptotic}), but in
the quenched case, were reported in refs.~\cite{bg,sharpe92}.

\section{Summary\label{sec6}}

In this work we computed the one-loop chiral corrections to the decay constant $f_K$ and to the bag parameter 
$B_K$ in all three versions of ChPT, i.e., full, quenched, and partially quenched. 
After working out the formulae in both infinite and finite volumes, we were able to discuss 
the impact of the finite volume effects on the chiral behavior of 
$f_K$ and $B_K$. We show that in most situations the physical chiral logarithms are 
completely drowned in the finite volume artefacts. In other words, unambiguously disentangling  
the physical chiral logarithms from the finite volume lattice artefacts does not appear to be feasible 
unless very large volumes are used.

We also discussed the shift of $f_K$ and $B_K$ induced by the finiteness of the volume. 
In our discussion we fix the strange quark in the kaon to its physical mass $m_s$, whereas the mass 
of the light quark is varied between $m_s/25$ and $m_s$. This mimics the current lattice 
QCD studies in which the strange quark 
is directly accessed on the lattice while the accessible light quarks have $r=m_q/m_s \in (0.5,1)$, so that 
an extrapolation to the physical $r_{u/d}=0.04$ is necessary.  

The results of our calculation indicate that for $r \gtrsim 0.25$ the finite volume effects are very small 
as long as $L \gtrsim 2$~fm. In that region we provide a simple asymptotic formula 
which is an accurate approximation of the full ChPT expressions.

From our formula it is also obvious that the finite volume corrections to $f_K$ and $B_K$ are different in quenched and
partially quenched QCD from those obtained in full QCD. Therefore, if in practical numerical simulations one 
wants to see only the effects of (un)quenching, the finite volume effects must be kept under control.

\vspace*{1cm}

\section*{Acknowledgment}

It is a pleasure to thank Gilberto Colangelo, Stephan D\"urr, Vittorio Lubicz and Guido Martinelli for discussions 
and comments on the manuscript. Partial support by the EC's contract HPRN-CT-2000-00145 
``Hadron Phenomenology from Lattice QCD" is kindly acknowledged.

\vspace*{1.6cm}

%%%%%%%%%%%%%  References

\end{document}